\documentclass[letter,english]{article}
\usepackage{emulateapj,onecolfloat,epsfig}

\begin{document}
\newcommand {\ds}{\displaystyle}
\def\sun{\hbox{$\odot$}}

\twocolumn
[
\title{Weak Lensing as a Calibrator of the Cluster Mass-Temperature Relation}
\author{Dragan Huterer$^1$ and Martin White$^2$}
\affil{$^1$Department of Physics, Case Western Reserve University, 
	Cleveland, OH~~44106}
\affil{$^2$Departments of Astronomy and Physics, University of California, 
	 Berkeley, CA~~94720}	

\begin{abstract}
The abundance of clusters at the present epoch and weak gravitational
lensing shear both constrain roughly the same combination of the power
spectrum normalization $\sigma_8$ and matter energy density
$\Omega_M$. The cluster constraint further depends on the
normalization of the mass-temperature relation. Therefore, combining
the weak lensing and cluster abundance data can be used to accurately
calibrate the mass-temperature relation.  We discuss this
approach and illustrate it using data from recent surveys.
\end{abstract}
 ]

\section{Introduction}

The number density of galaxy clusters as a function of their mass, the mass
function, and its evolution can provide a powerful probe of models of
large-scale structure.
Historically the most important constraint coming from the present day
abundance of rich clusters has been the normalization of the {\it linear
theory\/} power spectrum of mass density perturbations (e.g.\
\cite{Evr89,FWED,BonMye91,HA91,Lil,OukBla,BahCen,WEF,VL96,VL98,Henry}).
The normalization is typically quoted in terms of $\sigma_8$, the rms density
contrast on scales $8\,h^{-1}\,$Mpc, with the abundance constraint forcing
models to a thin region in the $\Omega_M$-$\sigma_8$ plane.

Since the mass, suitably defined, of a cluster is not directly observable, one
typically measures the abundance of clusters as a function of some other
parameter which is used as a proxy for mass.
Several options exist, but much attention has been focused recently on the
X-ray temperature.
Cosmological N-body simulations and observations suggest that X-ray
temperature and mass are strongly correlated with little scatter
(\cite{EMN,BryNor,ENF,HorMS,NevMF}).
How well simulations agree with observational results is far from clear,
and several issues need to be resolved.  On the simulation side there are
the usual issues of numerical resolution and difficulties with including all
of the relevant physics.  On the observational side instrumental effects can
be important (especially for the older generation of X-ray facilities) in
addition to the worrying lack of a method for estimating ``the mass''.
In this respect it is worth noting that there are numerous differing
definitions of which ``M'' and ``T'' are to be related in the M--T relation
(\cite{White_mass})!

With current samples the {\it dominant\/} uncertainty in the normalization
in fact comes from the normalization of the M--T relation
(\cite{ECF96,VL96,DV99,Henry,PSW,Seljak}).
Or phrased another way, the cluster abundance is a sensitive probe of the
normalization of the M--T relation.

The abundance of clusters is, of course, not the only way to constrain the
cosmological parameters.
In this regard it is interesting to note that weak gravitational lensing 
provides a constraint on a very similar combination of $\Omega_M$ and
$\sigma_8$.
Therefore, the two constraints can be combined to check for consistency of
our cosmological model, to provide a normalization for the M--T relation,
to probe systematics in either method and/or to measure other parameters
not as yet included in the standard treatments.

While the cluster constraint comes primarily from scales of about
$R=10\,h^{-1}\,$Mpc, current weak lensing surveys constrain somewhat
smaller scales.  These surveys probe scales between roughly 1 and 10
arcmin, which for source galaxies located at $z\simeq 1$ in a
$\Lambda$CDM cosmology corresponds to $0.7\,h^{-1}\,{\rm
Mpc}<R<7\,h^{-1}\,{\rm Mpc}$.  Therefore, weak lensing probes
slightly smaller scales than clusters.  As lensing surveys push to larger
scales the overlap will become even better.

In this paper we argue that a natural application of combining the
cluster abundance and weak lensing constraints is to calibrate the
M--T relation for galaxy clusters (see also \cite{HuKrav}).  In
Sec.~\ref{sec:MT} we define the M--T relation and derive how cluster
abundance constraints depend on $\Omega_M$ and $\sigma_8$.  In
Sec.~\ref{sec:WL_clus} we illustrate how combining the two constraints
can fix the normalization of the M--T relation using two recently
obtained data sets.  Finally, in Sec.~\ref{sec:concl} we discuss this
approach further.

\section{The Mass-temperature relation}\label{sec:MT}

Throughout we shall be interested in the abundance of massive clusters at
low redshifts, so we parameterize the M--T relation as
\begin{equation}
  {M(T,z)\over M_{15}}
  = \left( {T\over T_*} \right)^{3/2}
    \left(\Delta_c E^2 \right)^{-1/2} 
    \left[ 1 -  2  {\Omega_{\Lambda}(z)\over\Delta_c  }\right]^{-3/2}
\label{eq:MT}
\end{equation}
where $M_{15}= 10^{15}\,h^{-1}\,{\rm M}_\odot$,
$\Delta_c$ is the mean overdensity inside the virial radius in units of the
critical density, which we compute using the spherical top-hat collapse
model, and $E^2=\Omega_M (1+z)^3 + \Omega_{\Lambda} + \Omega_{\rm k} (1+z)^2$.
$T_*$ is the normalization coefficient that we seek to constrain; it roughly
corresponds to the temperature of a
$M=7.5\times 10^{13}\,h^{-1}\,{\rm M}_\odot$ cluster.
If measured in keV, the value of $T_*$ is precisely equivalent to $\beta$
from Pierpaoli, Scott \& White (2002) and is $1.34f_T$ of 
Bryan \& Norman (1998).

Let us explore the sensitivity of cluster abundance on $\Omega_M$ and
$\sigma_8$.  The Press-Schechter formula gives the number of collapsed
objects $dn$ per mass interval $d\ln M$ (\cite{PS}); we define
${\cal N}(M,z)=dn/d\ln M$.
Further defining $\nu\equiv \delta_c/\sigma(M, z)$, where $\sigma(M, z)$
is the rms density  fluctuation on mass-scale $M$ evaluated at redshift $z$
using linear theory and $\delta_c\approx 1.686$ is the linear threshold
overdensity for collapse, we have
\begin{equation}
  {\cal N}(M,z)= \sqrt{\frac{2}{\pi}} \frac{\rho_M}{M} 
  \frac{d\ln\sigma(M, z)}{d\ln\nu}\,
  \nu\,\exp\left( -\nu^2/2\right)
\label{eq:dndm}
\end{equation}
where $\rho_M$ is the present-day matter density. Assuming 
we are dealing with the current cluster abundance, $z\simeq 0$.
Following Pen (\cite{Pen}), for the mass scales of interest we can
approximate $\sigma(M)\propto M^{-\alpha}$ where $\alpha\simeq 0.27$ for
the currently popular $\Lambda$CDM cosmology.

Let us examine the dependence of ${\cal N}$ on $\Omega_M$, $\sigma_8$ and
$M$.  Ignoring the term $d\ln\sigma/d\ln\nu$ (which slowly varies) one
obtains
\begin{equation}
  {\delta{\cal N}\over {\cal N}} = \frac{\delta\Omega_M}{\Omega_M}
  (1-\alpha+\nu^2\alpha)+ \frac{\delta\sigma_8}{\sigma_8}(\nu^2-1)
  -\frac{\delta M}{M}(1-\alpha+\nu^2\alpha)
\end{equation}
Setting the left-hand side to zero and using the fact that $\delta
M/M=-3/2 \:\delta T_*/T_*$, for our fiducial cosmology and massive
clusters ($M\sim 10^{15}\,h^{-1}M_\odot$, or $\nu\simeq 2$) we
have\footnote{Note that the dependence of $M$ (or $T_*$) on $\sigma_8$ is
stronger for more massive clusters; a more detailed analysis gives
$T_*\propto \sigma^{-5/3}$ for the most massive clusters (\cite{Evr02}).}
\begin{equation}
  T_* \propto (\sigma_8\Omega_M^{0.6})^{-1.1}.
\end{equation}

Therefore, measurements of the cluster abundance at the present epoch
constrain a degenerate combination of $T_*$ and $\sigma_8\,\Omega_M^{0.6}$.
One of them cannot be determined without knowing the other.
Thankfully, weak lensing happens to measure roughly this combination of
$\Omega_M$ and $\sigma_8$ accurately, and the orthogonal combination much
less accurately (e.g.~\cite{Ber}).
Consequently, weak lensing in conjunction with cluster abundance can be
used to constrain $T_*$ quite strongly.

\section{Weak lensing plus clusters: an example}\label{sec:WL_clus}

As a more concrete example of these ideas, let us examine what value of
$T_*$ is required to bring current cluster and weak lensing results into
agreement.  This analysis will necessarily be illustrative, but is already
quite enlightening.

\subsection{The cluster data}

We compute $\sigma_8$ using a Monte-Carlo method following the steps
outlined in Pierpaoli, Scott \& White (2002).  Since some of the
details have changed we sketch the procedure here.

We use the HiFluGCS cluster sample of Reiprich \& B{\" o}hringer
(1999), restricted to clusters with $0.03<z<0.10$.  For simplicity we
do not include `additional' clusters of lower flux/temperature which
could scatter into the sample.  The cosmic microwave background (CMB)
frame redshifts from Struble \& Rood (\cite{StrRoo}) were used when
available and so were the two-component temperatures published in
Ikebe et al.~(2002).  For each $\Omega_{\rm m}$ we sample from a
distribution of cosmological parameters including $h$, $n$ and $T_*$
(the normalization of the M--T relation).  For each such realization
we generate 50 mass functions, where the temperature is chosen from a
Gaussian with the mean and variance appropriate to the observational
value and errors, and a scatter of 15\% in mass at fixed $T$ is
assumed for the M--T relation.  Using the mean values of the M--T
relation and the L--T relation from Ikebe et al.~(2002)
\begin{equation}
  L_X = 1.38\times 10^{35}\ \left( {kT\over 1{\rm keV}} \right)^{2.5}
  \quad h^{-2}{\rm W}
\end{equation}
we compute the volume to which clusters of mass $M$ could be seen above
the flux limit
$f_{\rm lim}=1.99\times 10^{-14}\,{\rm erg}\,s^{-1}\,{\rm cm}^{-2}$
of the survey.
For each realization of the mass function we compute the best fitting
$\sigma_8$ by maximizing the Poisson likelihood of obtaining that set of
masses from the theory with all parameters except $\sigma_8$ fixed.
The mass function can be computed using either the Press-Schechter (1974),
Sheth-Tormen (\cite{SheTor}) or Jenkins et al.~(\cite{JFWCCEY}) formulae.
We have used the Sheth-Tormen prescription throughout, with the mass variance
$\sigma^2(M)$ computed using the transfer function fits of Eisenstein
\& Hu~(\cite{EisHu}) and masses converted from $M_{180\Omega}$ to
$M_{\Delta_c}$ assuming an NFW profile (\cite{NFW}) with $c=5$.
The best fitting $\sigma_8$ is corrected from $\bar{z}$ to $z=0$.
The mean of the 50, $z=0$ normalizations is then taken as the fit for
that set of cosmological parameters (since the error from Poisson sampling
is completely sub-dominant to the error in the M--T normalization we do
not keep track of it here).  When quoting a best fit for a given triplet
of ($\Omega_{\rm m}$, $\sigma_8$, $T_*$), we marginalize (average) over the
other cosmological parameters $h$ and $n$.

\subsection{The weak lensing data}

As an example of weak lensing measurements, we use shear measurements
obtained using Keck and William Herschel telescopes (\cite{Bacon}).
These joint measurements used two independent telescopes covering 0.6
and 1 square degrees respectively, and enabled careful assessment of
instrument-specific systematics.
The authors compute the shear correlation function, and compare with the
theoretical prediction.
Assuming the shape parameter $\Gamma=0.21$, the results are well fit by
\begin{equation}
  \sigma_8 \left ({\Omega_M\over 0.3}\right )^{0.68}=0.97\pm 0.13
\end{equation}
which captures the total 68\% CL error: statistical, redshift uncertainty
and uncertainty in the ellipticity-shear conversion factor.
These results are consistent with other recent measurements of cosmic shear
(\cite{vW02,Refregier,Hoekstra}).

\subsection{Calibrating the M-T relation}

\begin{figure}[t]
\epsfig{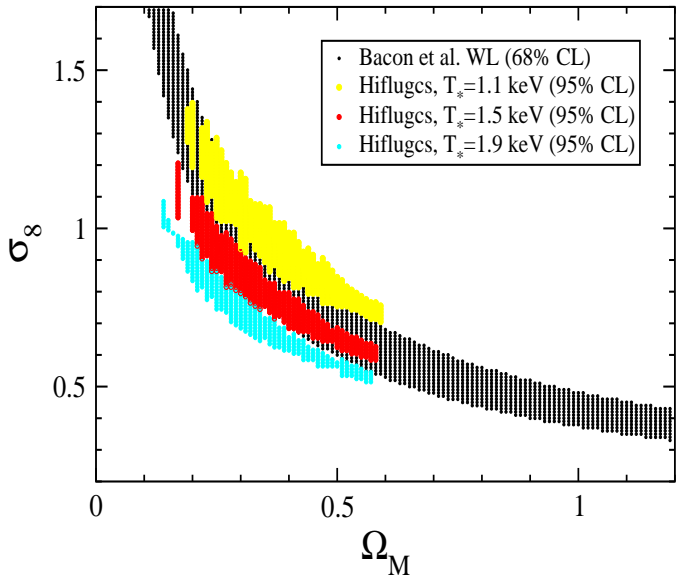}
\caption{68\% CL uncertainty contours in the $\Omega_M$--$\sigma_8$ plane, 
for a weak lensing survey (\cite{Bacon}) and 95\% CL uncertainties
for a cluster survey (\cite{ReiBoh}).  The cluster results are shown
for three different values of the mass-temperature normalization
parameter $T_*$, and marginalized over $n$ and $h$.  The degeneracy
regions for the two methods are very similar, which in principle
enables an accurate determination of $T_*$.}
\label{fig:WL_clus}
\end{figure}

Fig.~\ref{fig:WL_clus} shows the constraints in the $\Omega_M$-$\sigma_8$
plane.  The cluster constraint has been marginalized over $h$ and $n$ as
explained above, and plotted for three different values of $T_*$.
We have checked that the allowed ranges for $h$ and $n$ are wide enough so
that essentially all of the likelihood is contained within those ranges.
The weak lensing constraints assume the shape parameter $\Gamma=0.21$.
Note that the constraint regions from the two methods are indeed parallel,
with very similar degeneracy directions.
This enables an accurate determination of the normalization $T_*$.

In the example above, we see that a relatively low $T_*$ is preferred
($T_*\lesssim 1.7\,$keV) in order for cluster results to agree with
the weak lensing results.  While systematics in both methods could
still be important, it is interesting to note that this result is in line
with most earlier estimates (\cite{EMN,ENF,BryNor,YJS}), while it disagrees
with values adopted more recently (e.g.~\cite{Seljak}).

The fact that cluster abundance and weak lensing probe different
scales opens a possibility that one might be able to secure the
agreement between the two methods by varying the shape of the power
spectrum or the spectral index $n$ rather than the M--T normalization.
Unfortunately the constraints we have combined above have individually
been marginalized over $h$ and $n$.
Ideally, one would combine the cluster and weak lensing likelihood functions
and then marginalize over the relevant parameters to get the probability
distribution of $T_*$:
\begin{eqnarray}
  P(T_*) &=& \int \mathcal{L}_{\rm clus}(T_*, \Omega_M, \sigma_8, n, h)
  \nonumber \\ 
  &\times& \mathcal{L}_{\rm WL}(\Omega_M, \sigma_8, n, h)\,
	d\Omega_M\, d\sigma_8\,dn\,dh.
\end{eqnarray}
Then the results would be manifestly independent of the power spectrum
parameters.  We do not have the ability to perform such an analysis here.

Note, however, that the scales probed by lensing and clusters are
quite close, separated an order of magnitude at most.  For example, it
would require a spectral tilt of $n\sim 1.2$ to make the recently
obtained ``low'' normalization from cluster abundance ($\sigma_8\sim
0.6$) agree with the ``high'' normalization from weak lensing
($\sigma_8\sim 0.9$), and such a high value of $n$ is already
disfavored by recent CMB experiments (\cite{max,boom,DASI,CBI}).

\section{Conclusions}\label{sec:concl}

There has been a lot of discussion recently regarding the value of
cluster normalization $\sigma_8$.  While the ``old'' results favor
$\sigma_8\sim 1$ (\cite{VL98,PSW} and references therein), several new
cluster abundance analyses favor a significantly lower normalization
(\cite{ReiBoh,Bor01,Viana_Nichol,Seljak,Ikebe,SDSS}).  The lower
normalization is also favored by the combined analysis of 2dF Galaxy
Redshift Survey and CMB data (\cite{2dF}).  On the other hand, recent
weak lensing results (\cite{vW02,Bacon,Refregier,Hoekstra}) tend to
favor a higher value of $\sigma_8$. The cause is of this discrepancy
between various measurements has not been identified yet; one
candidate is larger than anticipated systematic errors in one or both
methods.  Another possibility is the bias in the relation between the
mass and the observable quantity --- temperature or luminosity ---
used to construct the abundance of clusters.

\vspace{0.2cm}
\begin{figure}[!ht]
\epsfig{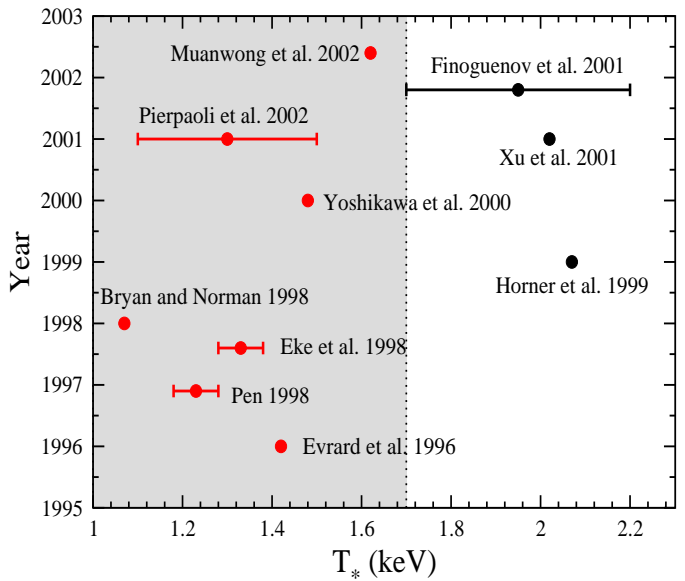}
\caption{Estimates of the M--T normalization $T_*$ 
collected from the literature. The red (light) points are estimates
from simulations, while the black (dark) points are from the
observations. Points with no error bars had none quoted. The shaded
region is roughly our favored range of values of $T_*$.}
\label{fig:MT_lit}
\end{figure}

The cluster abundance constraint on $\sigma_8$ crucially depends on
the M--T normalization $T_*$.  Figure~\ref{fig:MT_lit} summarizes the
current status of our knowledge of $T_*$. It shows seven
determinations from N-body simulations and three from direct
observations, as compiled in Pierpaoli, Scott \& White (2002) and Muanwong et
al.\ (2002). The shaded region is roughly our favored range of values
of $T_*$. Points without error bars had none quoted, and the three
observed values of $T_*$ assumed the isothermal-$\beta$
model. The measurement due to Muanwong et al.\ corresponds to their
``radiative'' and ``preheating'' cases that are cooling-flow
corrected, while the value due to Pierpaoli, Scott \& White is an
average over the simulations.  The large discrepancy between the
different measurements is apparent, and it also appears that the
observed values are systematically higher than the ones obtained from
simulations (see Muanwong et al.~2002 for further discussion).

We argue here that the cluster abundance -- weak lensing complementarity can
be used to cross-check the M--T relation.
By combining recent weak lensing constraints from Bacon et al.\ and the
HiFluGCS cluster sample of Reiprich \& B{\" o}hringer, we have demonstrated
the utility of this method.
While potential systematic errors in both data sets are still a concern,
the example we used prefers relatively low values of the M--T normalization
($T_*\lesssim 1.7\,$keV).
We conclude that future weak lensing surveys (Vista, LSST, SNAP) combined
with new cluster data from Chandra and XMM-Newton observations will provide
a strong probe of the M-T relation.

\section*{Acknowledgments}
DH is supported by the DOE grant to CWRU.  MW is supported by NASA and
by the Sloan Foundation.

\end{document}